\documentclass[12pt,preprint]{aastex}
\newcommand{\rxte}{{\it RXTE }}
\newcommand{\src}{4U 1820-30 }
\begin{document}

\title{On the nature of the flux variability during an expansion stage of 
a type I X-ray burst: Constraints on  Neutron Star Parameters for  4U 1820-30}

\author{Nickolai Shaposhnikov\altaffilmark{1}, 
        Lev Titarchuk\altaffilmark{2}}
\slugcomment{\it accepted by ApJL, February 23, 2004}

\altaffiltext{1}{George Mason University, School for Computational
Sciences; Center for Earth Observing and Space Research, Fairfax, VA 22030; 
nshaposh@scs.gmu.edu}
\altaffiltext{2}{George Mason University/Center for Earth
Observing and Space Research, Fairfax, VA 22030; and US Naval Research
Laboratory, Code 7620, Washington, DC 20375-5352; lev@xip.nrl.navy.mil }

\begin{abstract}
Powerful Type I X-ray burst with strong radial expansion  was 
observed from the low mass X-ray binary 4U 1820-30 
 with {\it Rossi X-ray Timing Explorer}  on May 2, 1997. 
We investigate closely the flux profile during the burst expansion stage.
Applying a semi-analytical model  we 
are able to uncover the behavior of a photospheric radius
and to simulate the evolution of neutron star (NS)-accretion disk
system.  We argue that although the bolometric luminosity is always  
the Eddington value $L_{Edd}$,  
 the photon flux at the bottom of the expanded envelope can decrease during the expansion stage. 
In fact, at the initial moment of explosion 
when the bottom burning temperature is  $\sim 2\times10^9$ K,  the bottom flux $L_{bot}$ is 
a few times  the Eddington limit, 
because the electron cross-section is a few times less than the Thomson cross-section 
at such a high temperatures. The surplus of energy flux with respect 
to the Eddington, $L_{bot}-L_{Edd}$,  
goes into the potential energy of the expanded envelope. As cooling 
of the burning zone starts the surplus decreases 
and thus the envelope shrinks while the emergent photon flux stays 
the same $L=L_{Edd}$. At a certain moment
the NS low-hemisphere, previously screened by the disk, becomes visible 
to the observer. Consequently, the  flux detected by  the
observer increases.  Indeed, we observe to the paradoxical situation 
when the  burning zone  cools,  but the apparent flux increases 
because of the  NS-accretion disk geometry. 
We demonstrate a strong observational evidence of NS-accretion disk occultation
in the behavior of the observed 
bolometric flux. We  estimate the anisotropy due to geometry and find 
that the 
system should have a high inclination angle.
Finally, we apply an analytical model of X-ray spectral formation  in the 
neutron star atmosphere during burst  decay stage to
 infer NS 
mass-radius relation.

\end{abstract}

\section{Introduction}
\src is one of the brightest low mass X-ray binaries (LMXB). It resides
in the globular cluster NGC 6624. Extremely short 685-second orbital period 
discovered by \citet{stella} from {\it EXOSAT} observations implied
that the system is ultra-compact and the secondary is a low-mass 
helium reach degenerate star. Using the analysis of UV 
diagrams of
NGC 6624 \citet[][ VLP hereafter]{vlp} evaluate the distance to be \src by $6.4\pm0.6$ kpc,
while the optical observations give a distance estimate of 7.6 kpc \citep{rml}. \rxte observations revealed \src as a 
prominent source of kilohertz quasi-periodic oscillations \citep{smale}.
Type I X-ray bursts \citep{lpt,str2003} from \src 
were discovered by \citep{gg}. It is now well established that these 
short ($\sim 10 - 15$ s) outbursts of energy
are due to the unstable thermonuclear burning of hydrogen/helium
mixture at the bottom of the NS atmosphere accumulated by accretion process.
SAS-3 observations showed strong evidence that 
X-ray bursts are seen only in its low-intensity  state \citep{clark}.
 All bursts observed from 4U 1820-30 show the radial expansion
with an apparent photospheric radius increase by a factor
of $\sim 20$. Such an expansion is accompanied by spectral 
softening which moves the spectrum completely out of
X-ray bandwidth and results in a so called 
``precursor'' effect \citep[see][and references therein]{str2003}.
 A several hour long ``superbursts''
was observed from \src on September 9, 1999. It is now believed 
that superbursts are
caused by the burning in the carbon ashes produced by
Type I bursts \citep{str2002}.

\citet{ehs83} were first to study hydrodynamics  
of an illuminated atmosphere of X-ray  burst at the peak (Eddington) flux.
Subsequently, the  X-ray spectral formation along with hydrodynamics during expansion 
and decay  stage of Type I bursts  were  studied  in \citet{t94a}, 
hereafter T94. Further development of the expansion stage solution is
presented in \citet{st2002}, hereafter ST02. 
In these papers the important effect of the radiative expansion was 
pointed out. Specifically, while the emergent photon luminosity is always  
the Eddington,  $L_{Edd}$, the photon flux at the bottom of
the expanded envelope can reach a value of a  few times greater
 than the Eddington limit. 
Indeed, at the initial moment of explosion 
when the bottom burning temperature is  $\sim2\times10^9$ K 
the bottom flux $L_{bot}$ is 
a few times more than Eddington luminosity, 
because the electron cross-section is a few times less than the Thomson 
cross-section 
for such a high temperatures (see below Eq. 1). 
The surplus of energy flux with respect 
to the Eddington limit, $L_{bot}-L_{Edd}$ 
goes into the potential energy of the expanded envelope. 
As cooling of the burning zone starts, the surplus decreases 
and thus the envelope shrinks (contraction stage of the radial expansion phase) while the emergent 
photon flux stays the same $L=L_{Edd}$. 
Shaposhnikov, Titarchuk \& Haberl (2003), hereafter STH,  draw attention to the 
fact that at the certain moment of the contraction stage
the NS low-hemisphere becomes visible to the observer, 
having previously been screened by the disk.
In this paper we provide more details into the physical insight of this 
 phenomenon and its relation to the data.

The spectral formation theory, employed in the present work,
 accounts for the effects of 
Comptonization, free-free absorption and emission.
Analytical expressions are derived 
for color factors, and the spectral shapes are presented as functions of 
the input parameters: atmospheric chemical composition, NS mass, radius
and the source distance. 
This technique was first applied to EXOSAT data from \src and 4U 1705-44
in \citet{ht}, hereafter HT95. In case of  4U 1820-30 data from 
seven bursts was used. For fixed valued of NS mass within the range 0.8-1.8 $M_\odot$
authors constrained NS radius using assumed source distance of 6.4$\pm0.6$ kpc.
In particular, for $d=6.4$ kpc and $M_{NS}=1.4 M_\odot$ they obtained the radius value of
7$\pm0.4$ km.
 In \citet{cygx2}, hereafter TS02 and STH where the technique is presented 
in detail, authors 
apply it to \rxte data from Cygnus X-2 and 4U 1728-34 respectively.  
  In this  Letter we  apply this
 methodology in analyzing the burst from 4U 
1820-30 observed by RXTE on May 2, 1997. High spectral and statistical 
quality of the data allows us to infer tight constrains on the NS
parameters. We find strong a
evidence of accretion disk-star geometry evolution,
that closely resembles the behavior of 4U 1728-34. 
 We compare our results for 4U 1820-30 with 
 equations of state (EOSs) of NS matter.


We present the Letter in the following manner. 
A brief description of the data  used in the analysis is given 
in \S 2. In  \S3 we consider the  general burst 
phenomenology. We specifically address its behavior during the 
 expansion stage in \S3.1 and calculate photospheric radius
and bottom temperature profiles. We discuss in detail
different processes which may affect the behavior of
the source flux during the burst expansion. We present the model 
and the results of its application to the burst data of \src
in \S 3.2. Specifically, we obtain the dependence of the NS mass on the radius 
as error contours, calculated for a set of distances to the system taken 
from the interval obtained by \citet{vlp}. 
We discuss  our results and come to conclusions in \S 4.

\section{Observations}
\src was observed by Proportional Array Counter \citep[PCA,][]{pca}
 on May 2, 1997 under Observation ID 20075-01-05-00. 
All five detectors were operative during the observation. 
In addition to the permanent Standard1 (one energy channel,
 1/8 sec time resolution) and Standard2 (129 energy channels, 16 sec time 
resolution) mode data, data in Event Mode with 125 $\mu$s time resolution and 
64 energy channels was recorded. Although the High resolution Event Mode 
overloaded the satellite telemetry  system 
during the peak of X-ray bursts, this
effect resulted in data loss only for three short
intervals during the expansion stage and thus did not
prevent us from being able to perform detailed spectral
analysis of the event. The burst started at  17:33:50 Terrestrial  Time. 
Prior the burst the source was in its low-intensity state with 3.3$\times 10^{-9}$ 
erg cm$^{-2}$s$^{-1}$ flux in 2-10 keV energy range.

Lightcurve of the burst is presented in \citet{str2002} where authors 
compared its general properties to the characteristics of superburst.
The apparent precursor  in the burst lightcurve 
is  an instrumental effect rather  than being a  real physical property of the burst.
It is due to the significant spectral softening during strong radial expansion.
Softening effect shifts the  spectrum emitted by 
the expanded NS photosphere completely out of the PCA energy bandwidth,
It is observed as a dip in PCA count rate. The initial rise of the 
photosphere is followed by 
the contraction stage  when the spectrum hardens and
 becomes detectable by {\it RXTE}/PCA detector array.
We start our spectral analysis of burst radiation beginning 2 seconds after
start of the burst when the spectrum shifts back to PCA detection bandwidth. 
   
\section{Data analysis and results}

\begin{figure}
\includegraphics[scale=1.0]{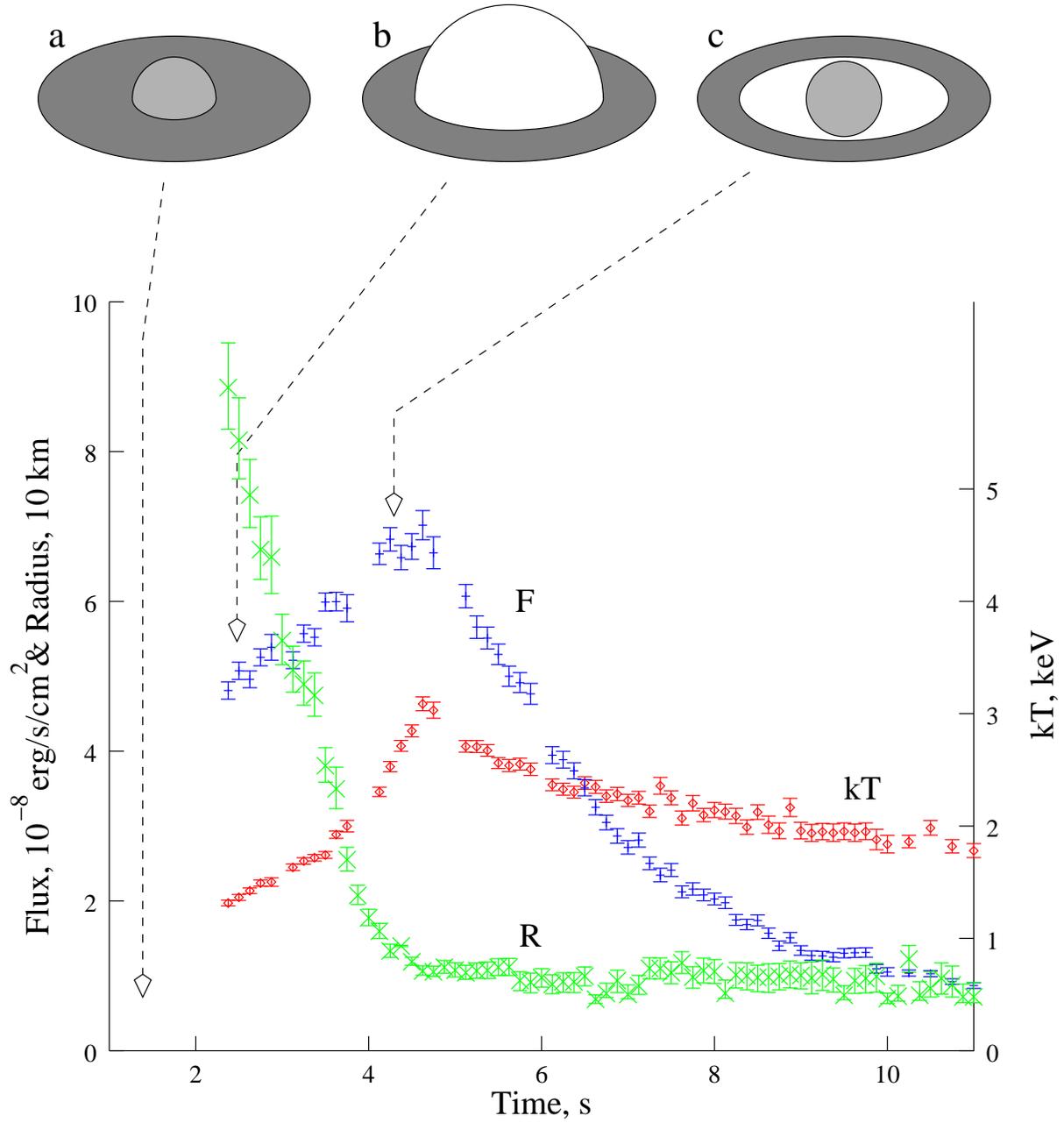}
\caption{Bolometric flux (red) and temperature (blue) of the
 burst spectrum versus time. Photospheric  radius given by the theory is shown in green.
The peaks of curves
correspond to the ``touchdown'' of the photosphere.  \label{flux_kT}}
\end{figure}

We use the standard method adopted for Type I X-ray burst
temporal spectral analysis. In the reduction of \rxte spectral data
we followed Recipes given in \rxte Cook Book.
We first extracted spectrum of the persistent
emission for 100 seconds immediately prior to the burst. We used
the extracted persistent spectrum as a background for the burst spectra.
During the burst we extract spectral slices for consecutive time 
intervals of 1/8 second. We used high resolution Event Mode for
extraction of both persistent and burst spectra and we utilized 
Standard1 mode for calculating all appropriate count rates for
deadtime corrections for the extracted spectra.
Counts from all three detector layers were added
during data reduction.
 We fitted burst spectra by the absorbed black body
shape. Quality of the fits is good for the entire decay stage
and for the most of the expansion episode with $\chi^2_{red}\sim 1.0$.
 We observe some excess in $\chi^2_{red}$  during particular episodes
of expansion episode when  $\chi^2_{red}$ goes up to 
$2.0-3.0$. It is presumably due to the high dynamics of the atmosphere.
This fact does not affect the results of our model fits 
for which we use only spectral data extracted from the burst decay stage. 
Temporal profiles of unabsorbed 0.001-100.0 keV model 
flux and temperature are given on Figure \ref{flux_kT}. 
The initial rise of the photosphere occurs in less than a 
second and the enormous expansion of NS atmosphere achieved within 
the first second of the burst. Subsequent contraction is accompanied
by spectral hardening and {\it bolometric flux growth}. 
Temperature and flux reach their  maxima of $3.1\pm0.1$ keV and 
$7.0\pm0.3\times 10^{-8}$ erg/(cm$^2$~s) simultaneously at the moment of photospheric ``touchdown''. 
After that both profiles decay exponentially. 

\subsection{Expansion stage}

The behavior of bolometric flux 
during the contraction of the burst
atmosphere is of special importance for our analysis.
 As it is shown on the magnified view  at the 
upper panel of Figure \ref{flux_kT}, observed flux gradually grows
as the photosphere contracts and color temperature increases. 
This effect can not
be explained by the gravitational redshift modulation  due to collapsing
radius. General relativistic (GR) effects 
 should, conversely, result in decreasing flux. 
It also can not be explained by 
relativistic corrections for Thompson scattering opacity \citep{pac}
expressed by
\begin{equation}
\kappa=\frac{\kappa_0}{1+(2.2\times T/10^{9}~{\rm K})^{0.86}}
\end{equation}
where $\kappa_0=0.2\,(2-Y_{He})$ cm$^2$/g. Through the part of the expansion episode during which the burst 
spectrum is observable by PCA the photosphere temperature rises 
from 1 keV up to 3 keV. The corresponding factor due to the 
scattering opacity relativistic correction is $\sim 1.06$ while 
the flux rises by the factor $\sim 1.35$.

Despite the fact that the amount of potential gravitational energy
deployed into uplifted layers of the atmosphere is considerable 
(see discussion below), the possibility
for its release in excess of the Eddintgton flux is highly questionable.
The release should occur at the bottom of the atmosphere and become
a part of radiation flux which is diffusing through the envelope and thus
is subjected to the Eddington limit.


With any  physical source of flux modulation being ruled out
we conclude that the system geometry should play a leading role
in determining flux behavior during photospheric contraction.
Namely, one should consider the
 dynamic evolution of ``NS-accretion disk'' system 
geometry to understand the flux time dependence.
The detailed discussion of the matter
is presented in STH, where the effect of the NS-disk 
occultation is reported for bursts with radial expansion from 4U 1728-34.
The proposed scenario of
NS-accretion disk interaction proceeds in the  following way. The expanded 
NS atmosphere effectively destroys the inner accretion disk
by evaporating it and pushing it away from the star surface.
The lower hemisphere of the photosphere, initially obscured from 
the observer by the disk, should now become visible. 
Radiation from the NS photosphere 
emerging from behind the disk inner edge results in the observed 
flux increase (see the top of  Fig. 1 for graphical representation 
of different geometry 
states of the system). In this picture it is 
implicitly assumed that the build up of the inner accretion disk
falls behind the photospheric contraction. This assumption is
reasonable, because the near-Eddington radiation pressure
should suppress the accretion until the touchdown, when
radiative flux drops lower $L_{Edd}$. Accretion disk comes
back and reaches NS surface after several second after touchdown,
 presumably during decay stage of the burst (see below).

Pure helium burning is characterized by a quick and powerful
energy release deeply  at the bottom of the atmosphere 
on a time-scale of $t_n\sim0.1$ sec \citep[for the details on nuclear burning 
on NS surface see review by][]{bs1998}.
The Eddington limit is exceeded and atmospheric expansion occurs.
The structure of the expanded atmosphere during the burst
is regulated by the  bottom conditions  (T94, ST02).
In fact, the observed radial expansion episode lasts about $3\sim5$ seconds.
 Thus the photospheric contraction is not governed by free fall , but
rather it is  quasi-steady, since cooling off the burning zone is 
relatively slow process, of order  seconds (see Spitkovsky et al. 2002).
  For the strong and extensive 
burst like in 4U 1820-30 the bottom temperature $kT_{b}$ is  
 $\sim 2\times10^9$ K and thus 
 the flux at the bottom is a few times larger than the Eddington limit because
of the attenuation of the cross-section due to the relativistic corrections 
(see Eq. 1). This fact allows the super-Eddington
energy rate to be effectively 
transfered into potential energy of the outer layers
of the atmosphere which are more opaque due to lower temperatures (T94).
Numerical results of ST02 show that overall energy
outflow from expanded envelope stays within 1-2\% of the Eddington radiation flux. 
In other words, the atmosphere acts as
a reservoir for  the surplus energy. It releases the stored energy at
the Eddington rate until photosphere reaches NS surface. 
According to the model atmosphere the potential energy of
the burst atmosphere during strong expansion can reach up to
$\sim10^{40}$ ergs. This energy supply is capable to explain the overall
energetics of the event.
Throughout
this particular contraction stage of radial expansion episode the 
bottom cools off while photospheric temperature
grows. The dynamic
of the system can be described by a sequence of a models related
to  various bottom temperatures.
All this picture is counterintuitive to the observer 
who  detects only the flux  at approximately the Eddington level and 
sees spectra that   harden 
during this stage.

\begin{figure}
\includegraphics[scale=0.65,angle=-90]{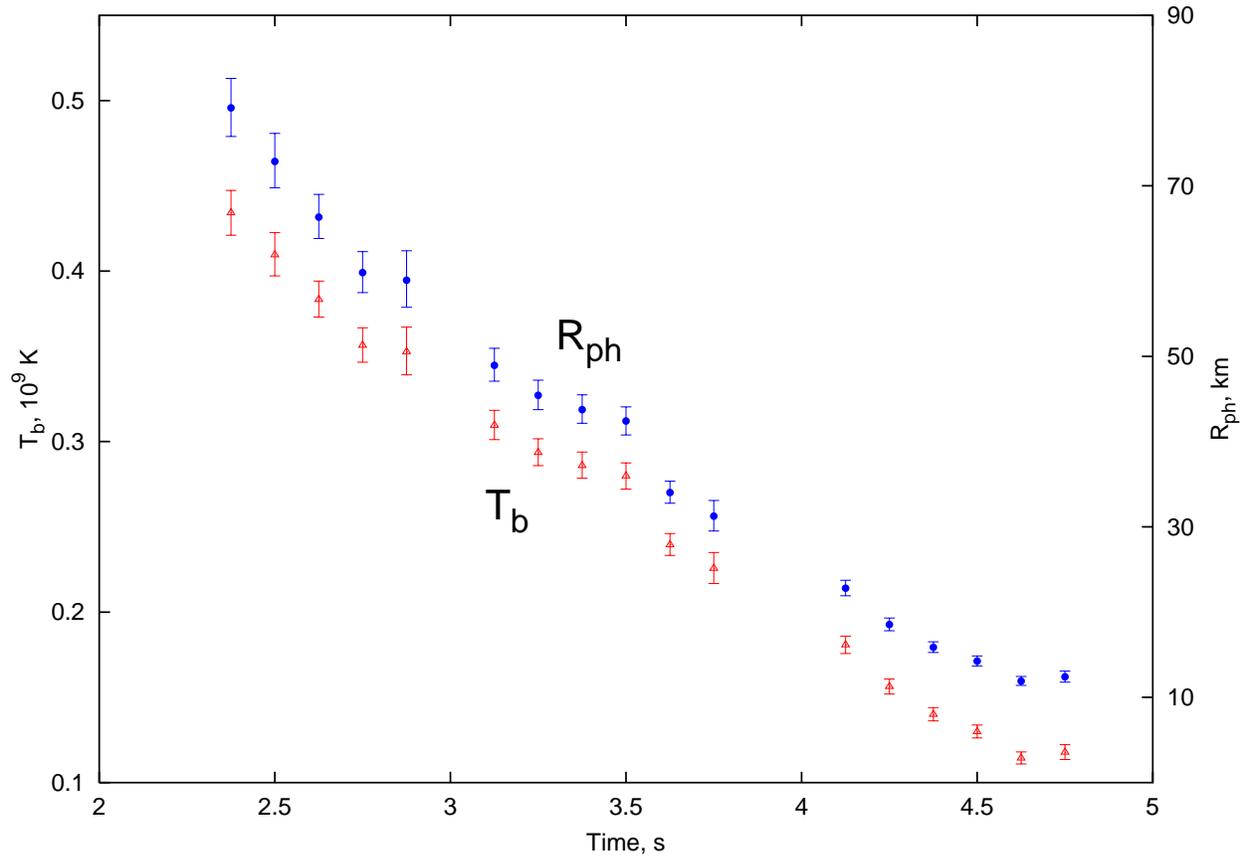}
\caption{Model bottom temperature (red) and photospheric radius (blue).
\label{tb_rph}}
\end{figure}

We investigate the behavior of the photospheric radius
applying semi-analytical theory developed by ST02.
Dependence of the various parameters of the expanded
envelope is presented as functions of the observed temperature of
the photosphere, its chemical composition and NS mass and radius.
Using pure helium atmosphere and NS parameters obtained in the next
Section for the source distance of 5.8 km ($m=1.29$, $R_{NS}=11.2$ km) we
 establish dependence of $R_{ph}$ on $kT$ and correspondingly on
flux. The dependence of $T_b$ on $R_{ph}$ is presented on
Figure \ref{tb_rph}. It is important to note that these
calculations are geometry independent because the color temperature  $kT$ is
a geometry independent quantity.  In fact, calculations of $M_{NS}$ and $R_{NS}$
(in the next section) are consistent with the  geometry evolution.
The hydrodynamic profiles, given by the model, allows us to 
evaluate the total mass $M_{env}$ of the expanded envelope. 
For the maximum temperature of $2\times10^9$ K and given NS
parameters  $M_{env}=5.8\times 10^{21}$ g. The corresponding
mass column of $M_{env}/4\pi R_{NS}^2 =3.6\times 10^8$ g/cm$^2$ is an agreement
with the column required for ignition (see  e.g. Bildsten 1998).

The apparent flux increase during contraction of the photosphere
is consistent with photospheric radius profile given by the
theoretical model. In total the  observed flux increase is a combination of three effects: 
increase due to geometry evolution, decrease due to gravitational 
redshift,  and slight increase due
to relativistic correction to opacity. The 
final expression for geometrical anisotropy is 
\begin{equation}
\label{anis}
\xi_b\approx 1.35(z+1)/1.06=1.26\, (z+1).
\end{equation} 
We include this equation in our model to account for
the anisotropy change during decay stage (see the next
section).

\subsection{Burst decay stage: NS mass and radius.}

We utilized  the theoretical model
for the  color temperature of the spectrum during the burst decay phase, i.e.
after touchdown moment 
(see T94). Here we present the
final formula for the color temperature $kT_{\infty}$ as a function
of input parameters of the problem
\begin{equation}
\label{eq1}
kT_{\infty}=2.1\, T_h\left[\frac{l\, m}{(2-Y_{He})(z+1)^3\,r_6^2}\right]^{1/4}
{\rm keV},
\end{equation}
where $m$ is the NS mass in units of solar mass, $r_6$ is the NS radius in 
units of 
10 km, $Y_{He}$ is a helium abundance, $l=L/L_{Edd}$ is the 
dimensionless luminosity in
 units of the Eddington luminosity. $T_h$ is the color 
(hardening) factor, which 
depends on $l$, and $Y_{He}$  (TS02). Parameters of the model are $m$,
$r_6$, $Y_{He}$ and $d_{10}$ is the distance to the object in units of 10 kiloparsec. 
The luminosity is expressed by 
\begin{equation}
l=0.476\,\xi_b\,d_{10}^2 F_8 (2-Y_{He}) (z+1)/m,
\end{equation}
where $\xi_b$  is anisotropy factor and $F_8=F/10^{-8}$ 
erg/cm$^2$s.  There are strong
indications that 4U 1820-30 is the pure helium accretor and bursts from
this source
originate in a helium-dominated environment \citep[see][]{lpt,str2003,bs1995,cumming}
so we put $Y_{He}=1.0$ for the entire model fitting procedure. 
The model gives the functional dependence of $kT_\infty$ upon
$F_8$ which are two observables obtained from spectral fits.
We fit our model to the data in the range $0.5 < F_8 < 6.5$, 
which approximately corresponds to $0.1 < l < 0.9$, using the 
lower limit to exclude the data points with large errors and 
possible systematic effects due to prominent, persistent component.
  We put  the upper 
limit on $l$ to distinguish safely from the fit the data points with  $l\sim 1$
 where expansion of the atmosphere can occur and the validity of the 
model becomes restricted.

First we assume that the anisotropy  of the system does not change
throughout burst decay. In this case  $\xi_b$ cannot be 
evaluated independently from distance because they come into model
only as a product $\xi_b\,d_{10}^2$. We put $\xi_b=1.0$ and 
perform the model fits by minimizing $\chi^2$.
 We  use three fixed values for the distance 5,8, 6,4 and 7.0 according
the error interval given by \citet{vlp}, and obtain  fits with
$\chi^2_{red}\geq 2.0$.
None of these fits are statistically  acceptable, if we require $\chi^2_{red} \sim 1.0$. 
 The model residuals have an apparent trend in
their deviation from the model. For $F_8 \lesssim2.5$ data points lie above
the model while for  $F_8 > 2.5$ they are below. The 
fit behavior for $\xi_b=$ const is consistent with the disk occultation
hypothesis mentioned in the previous section. According to this
hypothesis the entire NS is exposed to the observer   
until some moment, when accretion disk returns to NS surface.
We identify the time when the  disk comes back with the moment
 of the flux drop.
This occurs  around  seventh second of the burst
when the flux falls down to $\approx 2.5\times 10^{-8}$ erg/(cm$^2\cdot$s). 
Just before this moment the neutron star is open viewed and thus the system has the geometry corresponding
to $\xi_b=1.0$, i. e.
the disk subtends the lower NS hemisphere and the the system 
makes a transition to the  geometry state with $\xi_b^*$. 
 We modify 
our model to include the effect of the disk occultation  introducing
two additional parameters: anisotropy during occultation $\xi_b^*$
and dimensionless flux $F^*_8$ at the moment when occultation occurs.
In this analysis we consider only the dependence of $\xi_b^*$ upon the
inclination angle, i. e. $1.0 \leq \xi^*_b \leq 2.0$.
We use expression (\ref{anis}) for a self-consistent calculation of  $\xi_b^*$
during the model fitting.
When the occultation effect is consistently incorporated into the model
we obtain good quality fits with $\chi_{red}^2 < 1.0$. The results are summarized in Table 1.

\begin{figure} 
\includegraphics[width=5in,height=6in,angle=-90]{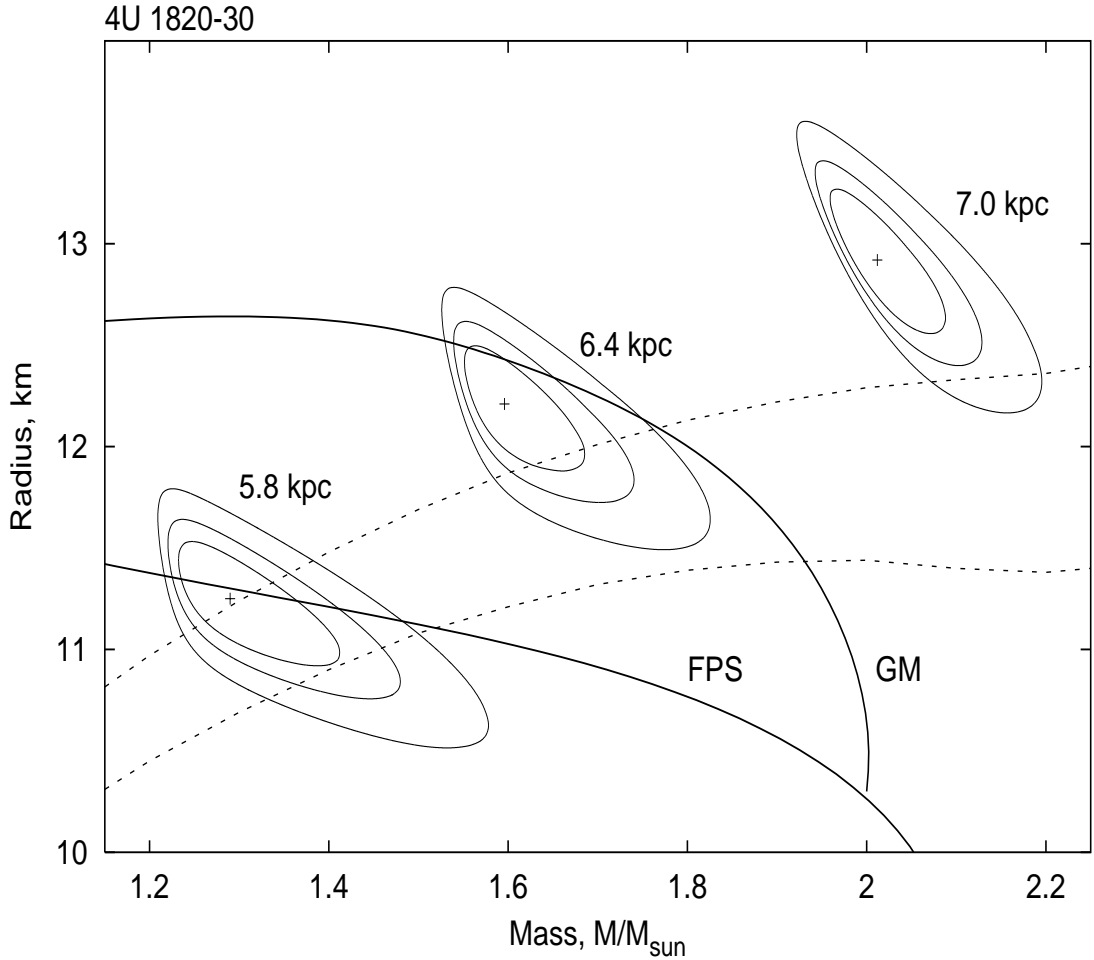}
\caption{ Mass-radius contour obtained by the model fitting. 
EOS mass-radius relations are presented by solid lines:
 FPS - \citet{fri}, GM - \citet{gm3}. Region, surrounded by dashed contour indicates the values
allowed by Eddington limit. 
\label{contours}}
\end{figure}

On Figure \ref{contours} we show the best-fit parameters and $M_{NS}-R_{NS}$ error contours for 68\%, 90\%  and 99\% confidence levels.
Curves GM and FPS presents M-R relationships 
for nuclear matter equations of state calculated by \citet{gm3} 
and \citet{fri} respectively. 
A reader who is interested in   this subject can find the details of  recent developments in NS EOS theory 
in \citet{ns_eos}. These particular details are  out of scope of our presentation.

The fact that during
expansion stage the Eddington limit is achieved  puts 
an additional constrain on the NS fundamental characteristics.
One  should observe gravitationally redshifted flux $F_{\rm Edd}$ from the unocculted
NS at the moment of the burst atmosphere touchdown, when
both flux and temperature peak (i.e on the fifth second on Figure \ref{flux_kT}).
 Clearly, the dimensionless luminosity  can be calculated as
$l=\xi_b F_8/F_{8, Edd}$.
Moreover, in this case the distance 
is not a parameter and we can calculate alternative domain 
in the M-R plane dictated by the observed Eddington flux.
 M-R values for 90\% confidence level, obtained with
this method, are shown by the dashed contour. M-R domains 
obtained with two methods depart from each other towards 
higher source distances. 
M-R domain is most consistent with distance of 5.8 kpc
while the 90\% contour obtained with two methods
for 7.0 kpc does not overlap.  This fact possibly  indicates 
lower values for the source distance and, correspondingly, 
NS mass as more favorable. We fail to find the model parameters
which describe the burst  temperature-flux dependence for 
the distance of 7.6 kpc given by optical measurements.

\section{Discussion and Conclusions}

In this {Letter} we report the results of application of
the theory of spectral formation during expansion and decay stage of
a Type I X-ray burst using  data from LMXB 4U~1820-30 collected with \rxte.

We investigate in detail the physics governing the 
burst properties during the expansion stage. 
The super-Eddington energy excess is deployed into
gravitational potential energy of the expanded atmosphere which 
is a result of the dependence of the electron opacity on
 temperature. The emergent luminosity (cooling rate of the expanded atmosphere) is at
the Eddington limit. When the bottom temperature drops
and photospheric contraction occurs,  one can intuitively expect 
the  overall flux decrease. However the exact contrary situation is 
observed. The photospheric temperature increases and
the bolometric flux grows presumably due to a geometrical NS
recovery from behind the accretion disk inner edge.
Simple arguments leads to the value of geometrically
induced anisotropy of  $\xi_b\approx1.55-1.71$ which
corresponds to the local system inclination angle 
$i\approx73^\circ- 80^\circ$ (see Table 1).

We infer the mass-radius relationship and  error
contours for a given distance by modeling the spectral 
temperature dependence on the bolometric flux.
 We include the effect of the radiation anisotropy due to change of the NS-disk geometry
into consideration and we estimate the anisotropy coefficient from
expansion stage behavior that is similar to 4U 1728-34 (STH).
Because of the superior instrumental capabilities of {\it RXTE} mission with 
respect to {\it EXOSAT} satellite,  
the data even for  one 
burst observed with {\it RXTE}  provides us more stringent constrains on the mass-radius relation 
 than several 
bursts from {\it EXOSAT}.
Both statistical behavior of
the model and Eddington flux limit strongly   
suggests that the distance to 4U 1820-30  close to 5.8 kpc, for which
we obtain $M_{NS}=1.29^{+0.19}_{-0.07}M_\odot$ and $R_{NS}=11.2^{+0.4}_{-0.5}$ km.


\clearpage 

\begin{deluxetable}{lcccccc}
\tablecolumns{8}
\tablewidth{0pt}
\tablecaption{Model Fits for 4U 1820-30\tablenotemark{a}.}
\tablehead{\colhead{$d$, kpc}&\colhead{$m$}&\colhead{$R_{NS}$, km}&\colhead{$\chi_{red}^2$}&\colhead{$z$}&\colhead{$i$}}
\startdata
5.8 &$1.29^{+0.19}_{-0.07}$&$11.2^{+0.4}_{-0.5}$& 0.74 & 0.23 & $73^\circ$\\
6.4 &$1.60^{+0.14}_{-0.06}$&$12.2^{+0.4}_{-0.5}$& 0.79 & 0.28 & $76^\circ$ \\
7.0 &$2.01^{+0.13}_{-0.07}$&$12.9\pm0.5$& 0.84 & 0.36 & $80^\circ$ \\
\enddata
\tablenotetext{a}{errors are given for 90\% of confidence}
\end{deluxetable}


\end{document}